\documentclass[11pt]{article}

\usepackage{graphicx}
\usepackage{amssymb}
\usepackage{amsthm}
\usepackage{subfigure}
\usepackage{url}
\usepackage{authblk}
\usepackage{amsmath}
\usepackage{color}

\usepackage{anysize}
\marginsize{1.5cm}{1.5cm}{1.5cm}{3cm}

\begin{document}

\title{Exploring Tehran with excitable medium}

\author[1]{Andrew Adamatzky}
\author[2]{Mohammad Mahdi Dehshibi}

\affil[1]{University of the West of England, Bristol, UK}
\affil[2]{Pattern Research Centre, Tehran, Iran}

\maketitle

\begin{abstract}
\noindent
An excitable chemical medium --- Belousov-Zhabotinsky (BZ) reaction --- is proven to be a fruitful substrate for prototyping unconventional computing devices. These include image processors, logical circuits, and robot controllers. We study a BZ potential for characterising a geometry of street networks on a fragment of Tehran street map. The city was chosen because it is one of the most populated cities in the World with nearly uncontrollable urban growth. In numerical experiments with Oregonator model of BZ reaction, we demonstrate that excitability of the medium allows acts as a selector between omnidirectional waves and soliton-like localised excitations. We uncover a phase-transition like dynamics, controlled by the excitability, of coverage of the street network by excitation wave-fronts. In the cluster analysis, we show how the network geometry, when it meets propagation of BZ wave-front, relates to the traffic flow of Tehran.\\
\vspace{5mm}
\end{abstract}

\section{Introduction}
\label{introduction}

A thin-layer BZ medium~\cite{belousov1959periodic, zhabotinsky1964periodic} shows the rich dynamics of excitation waves, including target waves, spiral waves and localised wave-fragments and their combinations. These waves can be used to explore the geometrical constraints of the medium's enclosure and to implement computation. An information processing, wet electronics and computing circuits prototyped in BZ medium include chemical diodes~\cite{DBLP:journals/ijuc/IgarashiG11}, Boolean gates~\cite{steinbock1996chemical, sielewiesiuk2001logical}, neuromorphic architectures~\cite{ gorecki2006information, gentili2012belousov, takigawa2011dendritic, stovold2012simulating,  gruenert2015understanding} and associative memory~\cite{stovold2016reaction,stovold2017associative}, wave-based counters~\cite{gorecki2003chemical},  arithmetic circuits~\cite{costello2011towards, sun2013multi, zhang2012towards, suncrossover, digitalcomparator}. Light sensitive modification, with Ru(bpy)$^{\text{+3}}_{\text 2}$ as a catalyst, allows for manipulation of the medium excitability and geometry of excitation wave fronts~\cite{kuhnert1986new, braune1993compound, manz2002excitation}. By controlling BZ medium excitability, we can produce related analogues of dendritic trees~\cite{takigawa2011dendritic}, polymorphic logical gates~\cite{adamatzky2011polymorphic} and logical circuits~\cite{stevens2012time}. 
We simulate light-sensitive BZ medium using  two-variable Oregonator model~\cite{field1974oscillations} adapted to a light-sensitive Belousov-Zhabotinsky (BZ) reaction with applied illumination~\cite{beato2003pulse}. 
The Oregonator equations are proven to adequately reflect the behaviour of real BZ media in laboratory conditions, including triggers of excitation waves in 3D ~\cite{azhand2014three}, phenomenology of excitation patterns in a medium with global negative feedback~\cite{vanag2000pattern}, controlling excitation with direct current fields~\cite{vsevcikova1996dynamics}, dispersion of periodic waves~\cite{dockery1988dispersion}, 3D scroll waves~\cite{winfree1989three}, excitation spiral breakup~\cite{taboada1994spiral}. Authors of the present paper employed the Oregonator model as a virtual test bed in designing BZ medium based computing devices which were implemented experimentally~\cite{adamatzky2007binary, de2009implementation, toth2009experimental, toth2010simple, adamatzky2011towards, stevens2012time}.  Therefore the Oregonator model is the ideal --- in terms of minimal description yet highest expressiveness --- computational substitute to laboratory experiments.

Exploration of space with oxidation waves fronts in BZ medium has been studied in the context of maze solving, shortest paths finding~\cite{agladze1997finding, steinbock1995navigating, rambidi1999finding}, and collision avoidance \cite{adamatzky2002collision}. These works employed a fully excitable medium, where a source of perturbation causes the formation of circular waves, and then wave-fronts propagate in all directions, and `flooding' all domains of the space. In sub-excitable BZ medium wave-fragments behave as dissipating solitons~\cite{hildebrand2001spatial, adamatzky2004collision, de2009implementation}, preserving their shape and velocity vector. Based on a success of our previous work on (sub-)excitable London streets~\cite{adamatzkyExcitableLondon}, we aimed to answer the following questions: (1) what elements of the Tehran street network would be preserved, in terms of being always spanned by travelling excitation, when excitability of the medium decreases, and, (2) how the propagation of excitation wave-fronts might relate to traffic flow in terms of changing excitability of the medium. We have chosen Tehran because a growth of the city, from its inception, was affected by a wide range of cultural, religious and political factors, which made their unique imprints on a geometry of Tehran street networks~\cite{kheirabadi2000iranian,brunn2003cities}. The city is amongst most populated cities in the world, suffering from traffic congestion and environmental pollution~\cite{madanipour2006urban}, with many areas having a high vulnerability to earthquakes~\cite{shieh2014earthquake, modarres2002application}, exemplifying social division and environmental risks~\cite{madanipour2011sustainable}.

\section{Methods}

\begin{figure}[!tbp] 
    \centering
    \includegraphics[width=\textwidth]{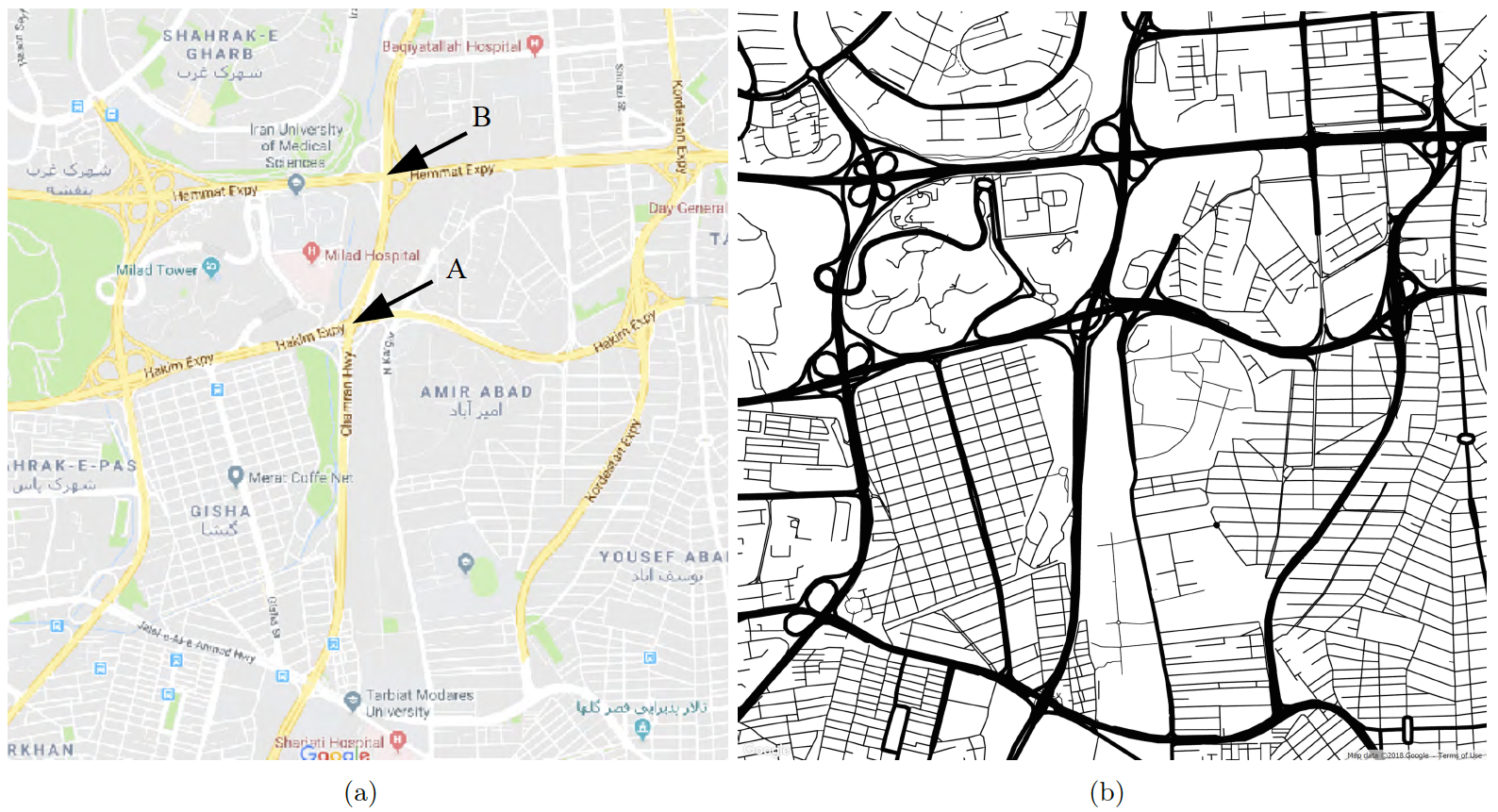}
    \caption{Fragment of Tehran street map used in computational experiments.  
(a)~Google map. Map data \copyright 2018 Google. Sites of initial perturbation is shown by arrow and labelled `A' and `B'. (b)~Template used for studies.}
    \label{fig:originalmap}
\end{figure}

A fragment of Tehran street map (map data \copyright 2018 Google) approximately 
3.9 km by 4.2 km, was mapped onto a grid of 2500 by 2500- nodes (Figure~\ref{fig:originalmap}). Nodes of the grid corresponding to streets are considered to be filled with a Belousov-Zhabotinsky medium, i.e. excitable nodes, other nodes are non-excitable (Dirichlet boundary conditions, where the value of variables are fixed zero). We use two-variable Oregonator equations~\cite{field1974oscillations} adapted to a light-sensitive 
Belousov-Zhabotinsky (BZ) reaction with applied illumination~\cite{beato2003pulse}:

\begin{eqnarray}
  \frac{\partial u}{\partial t} & = & \frac{1}{\epsilon} (u - u^2 - (f v + \phi)\frac{u-q}{u+q}) + D_u \nabla^2 u \nonumber \\
  \frac{\partial v}{\partial t} & = & u - v 
\label{equ:oregonator}
\end{eqnarray}

The variables $u$ and $v$ represent local concentrations of an activator, or an excitatory component of BZ system, and an inhibitor, or a refractory component. Parameter $\epsilon$ sets up a ratio of the time scale of variables $u$ and $v$, $q$ is a scaling parameter depending on rates of activation/propagation and inhibition, $f$ is a stoichiometric coefficient. 
Constant $\phi$ is a rate of inhibitor production. In a light-sensitive BZ, $\phi$ represents the rate of inhibitor production proportional to the intensity of illumination. The parameter $\phi$ characterises excitability of the simulated medium. The larger $\phi$ the less excitable medium is. We integrated the system using Euler method with five-node Laplace operator, time step $\Delta t=0.001$ and grid point spacing $\Delta x = 0.25$, $\epsilon=0.02$, $f=1.4$, $q=0.002$. We varied value of $\phi$ from the interval $\Phi=[0.05,0.08]$. The model has been verified by us in experimental laboratory studies of BZ system, and  the sufficiently satisfactory match between the model and the experiments was demonstrated in \cite{adamatzky2007binary, de2009implementation, toth2010simple, adamatzky2011towards}. 

To generate excitation wave-fragments we perturb the medium by square solid domains of excitation, $20 \times 20$ sites in state $u=1.0$, site of the perturbation is shown by the arrow in Figure~\ref{fig:originalmap}a. Time-lapse snapshots provided in the paper were recorded at every 150\textsuperscript{th} time step, we display sites with $u >0.04$; videos supplementing figures were produced by saving a frame of the simulation every 50\textsuperscript{th} step of numerical integration and assembling them in the video with play rate 30 fps.  All figures in this paper show time lapsed snapshots of waves, initiated just once from a single source of stimulation; these are not trains of waves following each other.

For chosen values of $\phi$, we recorded integral dynamics and calculated coverage of the streets network by travelling patterns of excitation. Integral dynamics of excitation calculated as a number of grid nodes with $u>0.1$ at each time step of integration.  A value of coverage is calculated as a ratio of nodes, representing streets, excited ($u>0.1$) at least once during the medium's evolution to a total number of nodes representing streets.

\section{Results}

To answer the  questions posed in Sect.~\ref{introduction}, we undertook series of numerical experiments and cluster analysis as following.

\subsection{Numerical Analysis}

\begin{figure}[!tbp] 
    \centering
     \includegraphics[width=\textwidth]{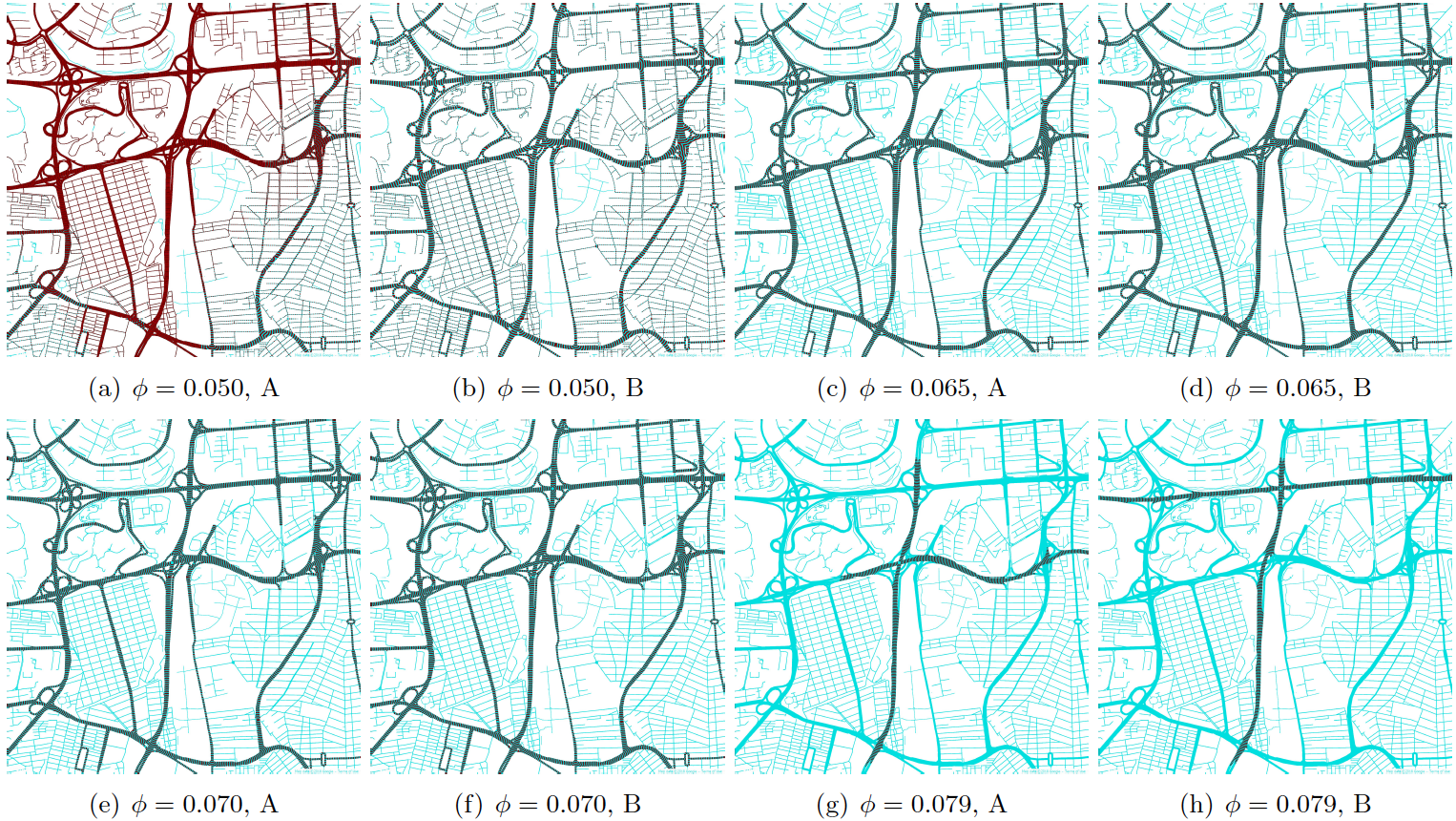}
    \caption{Propagation of excitation on the street map. Values of $\phi$ and perturbation sites are indicated in the sub-figure captions. Perturbation sites 'A' or `B' are indicated in Figure~\ref{fig:originalmap}a. These are time lapsed snapshots of a single wave-fragment recorded every 150\textsuperscript{th} step of numerical integration. Maps are generated using Processing www.processing.org. Videos, snapshots and data files are available at DOI 10.5281/zenodo.1304036. }
    \label{fig:lapses}
\end{figure}

\begin{figure}[!tbp] 
    \centering
      \includegraphics[width=\textwidth]{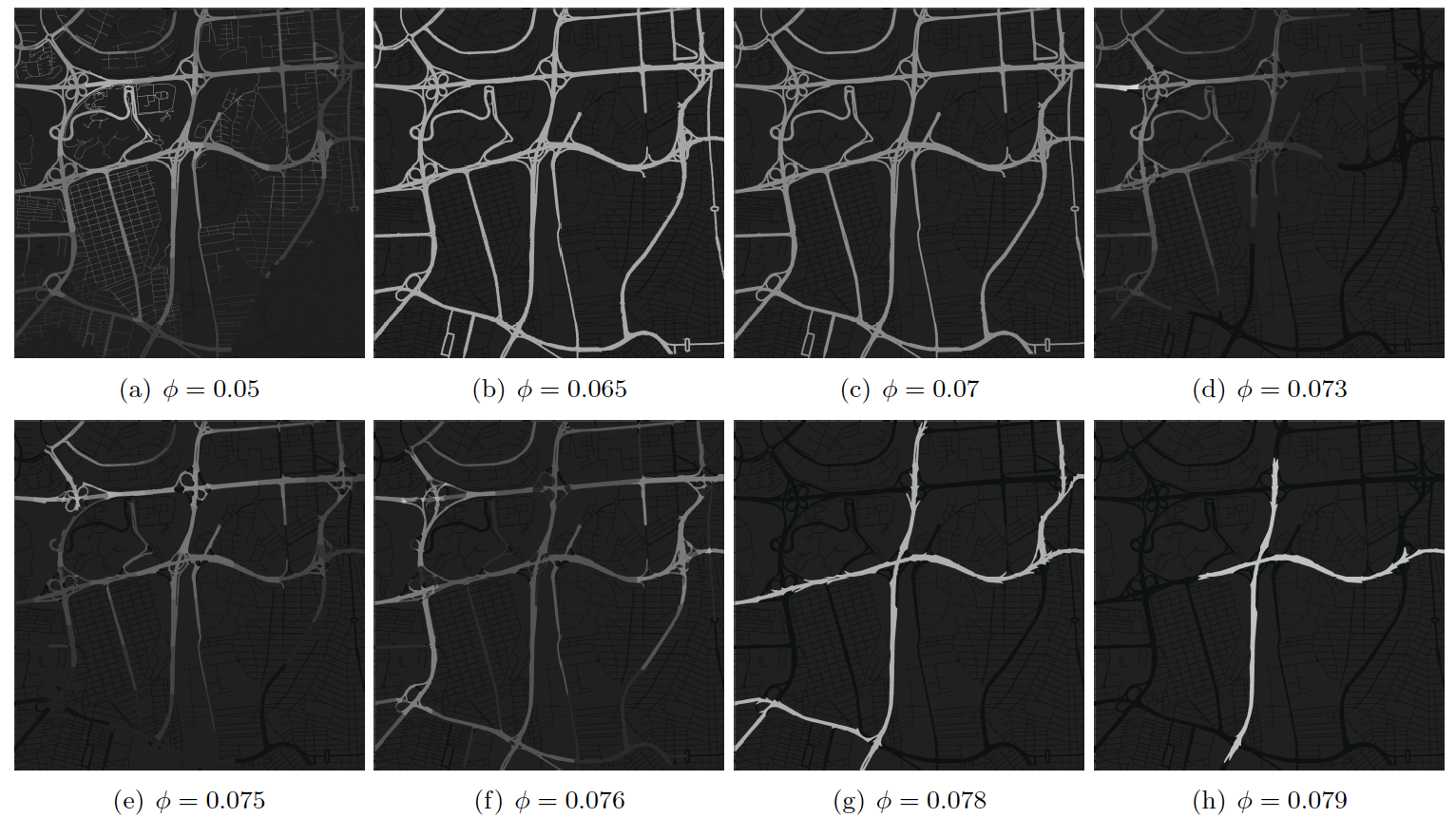}
    \caption{Coverage frequency visualisation. A brightness of a pixel is proportional to a number of times the pixel was excited normalised by a total number of excited pixels. Site A was perturbed.  Maps are generated using Processing www.processing.org.  Videos, snapshots and data files are available at DOI 10.5281/zenodo.1304036.}
    \label{fig:coveragefrequency}
\end{figure}

\begin{figure}[!tbp] 
    \centering
     \includegraphics[width=\textwidth]{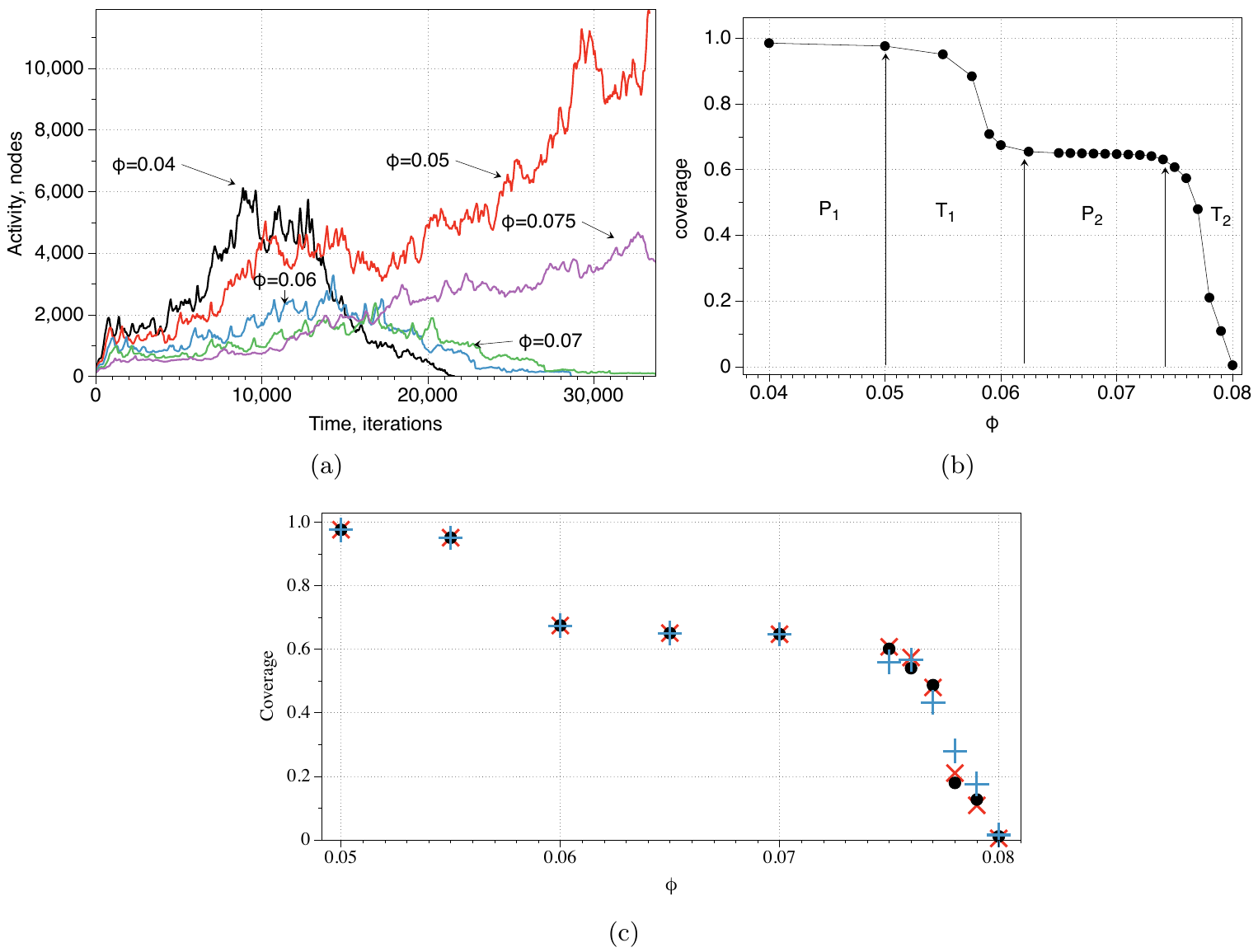}
    \caption{Integral characterisation of excitations on street networks. 
    (a)~Examples of activity, site A perturbed.
    (b)~Coverage for fine-grained range of $\phi$, site A was perturbed originally.
    (c)~Coverage for selected values of $\phi$ and perturbation sites A, $\times$-shape, B, $\bullet$-shape, and A and B at the same time, $+$-shape. 
     Videos, snapshots and data files are available at DOI 10.5281/zenodo.1304036.
    }
    \label{fig:plots}
\end{figure}

When a concentration of activator $u$ in the perturbation domain of 10 by 10 nodes is set to 1. Excitation wave-front is formed. The front expands, excitation enters streets branching out of the perturbation site, propagates along the streets and branches out in other streets, depending on excitation parameter $\phi$ (Figure~\ref{fig:lapses}). With the increase of $\phi$ from 0.05 to 0.08 less excitation propagates along fewer streets. This can be visualised using coverage frequency as shown in Figure~\ref{fig:coveragefrequency}.  Integral activity, i.e. a number of nodes excited at each step of the simulation, reflects space-time patterns of wave-fronts. In fully excitable regimes, $\phi=0.04$ and $\phi=0.05$ in Figure~\ref{fig:plots}a, we observe nearly exponential growth of activity --- while excitation wave-fronts are repeatedly branching at the street junctions and major part of the street network got traversed by the wave-fronts, see Figure~\ref{fig:lapses}ab and Figure~\ref{fig:coveragefrequency}a. The explosive growth of excitation abruptly comes to the halt when excitation wave-fronts reach absorbing boundaries of the simulated domain. With increase of $\phi$ to 0.06 the excitation activity shows lesser amplitude and extinct earlier, typically after 30K steps of integration, $\phi=0.06$ and $\phi=0.07$ in Figure~\ref{fig:plots}a. For $\phi=0.065$ to $0.76$ the excitable street network shows patterns of periodic activity, where excitation wave-fronts repeatedly appear along the streets, due to the excitation cycling along some paths. For this values of $\phi$ the integral activity never recedes but becomes sustained around some critical value (Figure~\ref{fig:plots}a, $\phi=0.075$).

Integral coverage, i.e. a ratio of nodes excited at some stage of the evolution to a total number of nodes, for an excitation initiated at site A is shown in Figure~\ref{fig:plots}b and $\times$-shapes in Figure~\ref{fig:plots}c. 
For several several values of $\phi$ the coverage was calculated for excitations initiated at site B, $\bullet$-shapes in Figure~\ref{fig:plots}c, and both sites A and B simultaneously, $+$-shape. Plot on Figure~\ref{fig:plots}c demonstrates that coverage is independent on a perturbation site, with nearly perfect match for sites A and B, therefore further we will deal with site A scenario.  The cover vs. $\phi$ plot consists of three phases $P_1$, $\phi \in ]0.04, 0.05]$,  
$P_2$, $\phi \in ]0.0625, 0.074]$,  $P_3$, $\phi >0.08$, and two phase transitions $T_1$, $\phi \in ]0.05, 0.0624]$, and $T_2$, $\phi \in ]0.075, 0.08]$  (Figure~\ref{fig:plots}b). In $P_1$ the medium is fully excitable and excitation wave-fronts propagate to all streets (Figure~\ref{fig:lapses}ab and Figure~\ref{fig:coveragefrequency}a), coverage is nearly 1. In $P_2$ excitation wave-fronts do not enter narrow streets, esp. branching out of larger street at nearly 90$^o$ (Figure~\ref{fig:lapses}cde and Figure~\ref{fig:coveragefrequency}bc); the coverage of the street network in this phase is c. 0.65.  In $P_3$ the medium becomes non-excitable. During transition $T_1$ coverage drops by third, most dramatic drop is observed in $T_2$ with coverage being a function $\phi$ as $10.689 + (-133.66)\cdot \phi$.

\subsection{Cluster Analysis}

\begin{figure}[!tbp] 
    \centering
     \includegraphics[width=\textwidth]{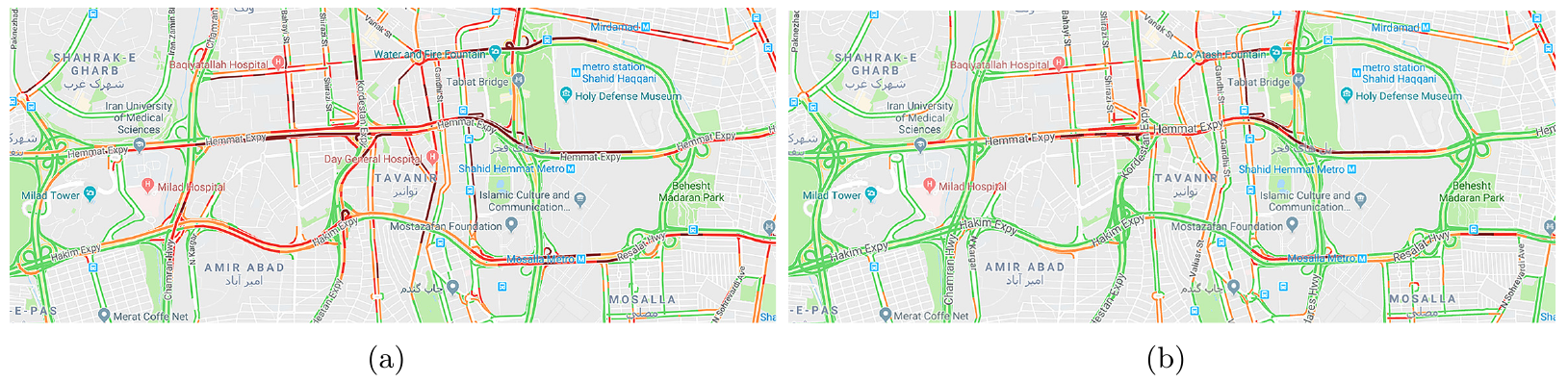}
    \caption{(a)~Live Tehran traffic map of Saturday. (b)~Live Tehran traffic map of Wednesday. Google live traffic.  Map data \copyright 2018 Google. Videos of live traffic time lapses are available at Zenodo DOI 
    10.5281/zenodo.1306936. }
    \label{fig:5}
\end{figure}

To uncover how the propagation of excitation wave-fronts might relate to traffic, we captured the live traffic of the selected district (Figure~\ref{fig:5}) during a week (25 May 2018 -- 1 June 2018). Snapshot from Google maps were captured to form the traffic time-lapse of the day. We observed that the speed of excitation wave-fronts has a direct relation to traffic propagation. For example, when the time-lapse of Saturday (Figure~\ref{fig:5}a) or Wednesday (Figure~\ref{fig:5}b) is compared with generated excitation wave-fragments --- $\epsilon = 0.02$, $\phi = 0.076$ and $\phi = 0.078$, respectively --- one can see that the shortest path algorithm would not suggest the optimal way to  navigate. Even on Wednesday, we are witnessing critical traffic conditions which Google map shows dark red in all hours in Hemmat highway and, as we see in Figs.~\ref{fig:lapses} and \ref{fig:coveragefrequency}, no excitation wave-fronts propagate in this highway; also compare of a full coverage (Figure~\ref{fig:coveragefrequency}b) and Tehran's traffic flow (Figure~\ref{fig:5}) in which Tehran transportation network experiences a deadlock traffic condition in almost all working days between 5.30 and 6.30pm.

\begin{figure}[!tbp] 
    \centering
      \includegraphics[width=0.7\textwidth]{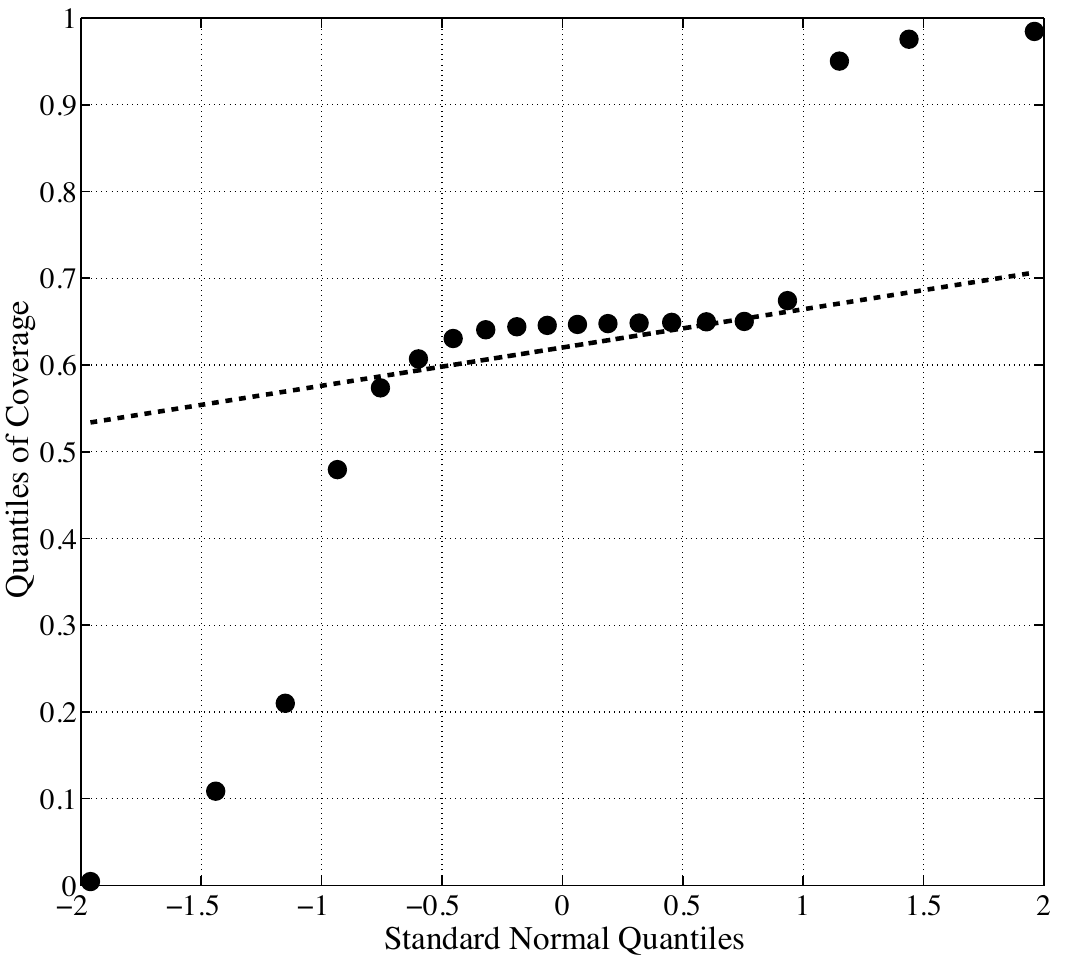}
    \caption{Q--Q plot for coverage values of of Tehran's street network by excitation waves, versus a normal distribution. The points follow a  nonlinear pattern, suggesting that the data are not distributed as a standard normal ($X \sim N(0,1)$). The offset between the line and the points suggests that the mean of the data is not 0 and points in the upper right corner of plot are those in which the traffic flow shows a chaotic behaviour.}
    \label{fig:6}
\end{figure}

We undertook a cluster analysis on the spatial coverage and super-positions. In the previous work \cite{adamatzky2018excitable}, we stated that the highest rate of coverage could reveal that increase of Reynolds number leads to the effect of street pruning and the coverage of streets by excitation waves is substantially different from that by fluid flow. However, in this study, we could find which $\phi$ is more compatible with the chaotic nature of Tehran traffic flow. Figure~\ref{fig:TehranLondon}a (Tehran) shows that the spatial coverage of the transportation network when it is spanned by excitation wave-fronts for different values of $\phi$ inversely relates to the medium's excitability (increasing of $\phi$). Then, we calculate the cumulative probability distribution functions of the spatial coverage, and its associated quantile function to compare it with a normal probability distribution by plotting their quantiles against each other (Figure~\ref{fig:6}). This plot, also known as quantile-quantile plot, helps us to compare if the empirical set of spatial coverage comes from a population with a normal distribution \cite{chambers2017graphical}. Let $F$ and $G$ be the cumulative probability distribution functions (CDF) of spatial coverages and a normal distribution, respectively. The inverse of CDF functions, $F^{?1}$ and $G^{?1}$, is the quantile function. The Q--Q plot draws the $q$\textsuperscript{th} quantile of $F$ against the $q$\textsuperscript{th} quantile of $G$ for a range of values of $q$. This plot selects quantiles based on the number of values in the sample data, i.e., if the sample data contains $n$ values, then the plot uses $n$ quantiles in which the $i$\textsuperscript{th} ordered statistic is plotted against the $\frac{i-0.5}{n}$\textsuperscript{th} quantile of the normal distribution, $X \sim N(0,1)$.


 \begin{figure}[!tbp] 
    \centering
      \includegraphics[width=\textwidth]{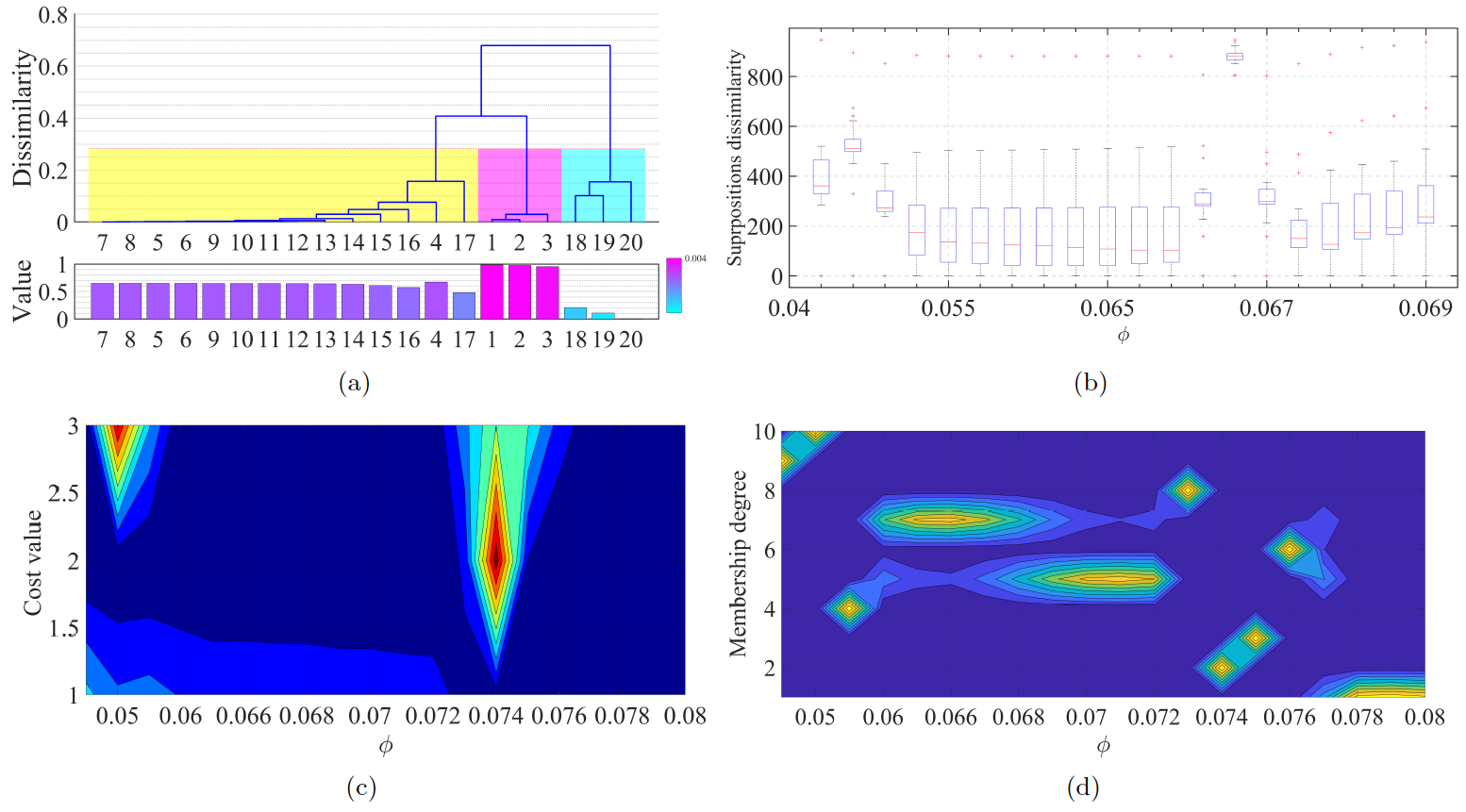}
    \caption{
    (a)~Dendrogram of hierarchical clustering for spatial coverage values of of Tehran's street network. The points are subject to calculating dissimilarity matrix by Euclidean distance. The points on horizontal axis do not follow a regular flow, even within a cluster. In terms of corresponding values to each $\phi$ within a cluster, one could observe a monotonic decrease, except for $\phi = 0.060$.
    (a)~Dissimilarity of super-positions calculated using Euclidean distance. 
    (b)~Cost of clustering super-position data using PSO when the number of clusters, $k =10$.
    (c)~Membership degree versus $\phi$ for clustering super-position using FCM into $k=10$ clusters. 
    }
    \label{fig:10}
\end{figure}

The plotted points in the Q--Q plot are non-decreasing when viewed from left to right. As the general trend of the Q--Q plot is flatter than the line $y = x$, the plot of a normal distribution is more dispersed than the distribution of spatial coverage rates. The ``S" shape of coverage distribution indicates that it is more skewed than the normal distribution. Coverage values related to the $\phi$ in the range of 0.065--0.077 are fallen on the line which is matched to our recent observation that the traffic flow does not follow the shortest path algorithm in a standard navigation system. Moreover, when $\phi = 0.078$, which is in accordance to chaotic nature of Tehran traffic, falls in the tails of Q-Q plot, it can reveal that the coverage distribution has a heavier weight than a normal distribution does. Hierarchical clustering of spatial coverage shows that three clusters could be obtained. Figure~\ref{fig:10}a shows the dendrogram of the conducted experiment. The clustering results accredit Q-Q plot, where the points on the tails of the graph are put in the same clusters. Therefore, it is reasonable to divide the experiments range into two sub-ranges, where $R_{1} = \{\phi | \phi \in [0.040, 0.080]\}$ and $R_{2} = \{\phi|\phi \in [0.060, 0.077]$. In Figure ~\ref{fig:10}a, one can see a behaviour in the cluster of $R_{2}$ for $\phi = 0.060$.

Finally, by calculating dissimilarity of super-positions of data acquired from the numerical integration of the model, and clustering with Fuzzy C-means and PSO-based clustering~\cite{yazdani2012multilevel, dehshibi2017hybrid}, we demonstrate two phenomenological discoveries related  to Tehran traffic flow:
\begin{itemize}
    \item  Dissimilarity of super-positions in $\phi= 0.050$ and $\phi= 0.074$ grows substantially (Figure~\ref{fig:10}b). When this data is clustered using particle swarm optimisation (PSO), the cost function shows a similar behaviour (see Figure~\ref{fig:10}c). This means that the $\phi$ values are proportional to the starting and finishing hours of Tehran congested traffic condition which is previously discussed based on $\zeta$.
    \item When super-position data is clustered using Fuzzy C-means (Figure~\ref{fig:10}d), four clusters are observable. Based on the degree of membership, $\phi = 0.076$ is an isolated cluster, which is similar to a deadlock traffic behaviour in Tehran street in which any commuting on Hemmat highway is almost impossible and we observed that no excitation waves are propagating in this area of Tehran street network. Points with $\phi$ values of 0.074 and 0.075 corresponds to heavy traffic conditions. For the $\phi \in [0.055, 0.073]$ we observe two semi-overlapped clusters which represent moderate traffic conditions. There are two clusters: the first cluster contains $\phi \in [0.040, 0.050]$ and the second cluster has $\phi \in [0.078, 0.080]$. These clusters contain super-positions of BZ propagated over Tehran street network where free or moving traffic was recorded. These are similar to conditions of traffic flow leaving or entering the state of moderate traffic. 
\end{itemize}

\section{Discussion}

\begin{figure} 
    \centering
      \includegraphics[width=\textwidth]{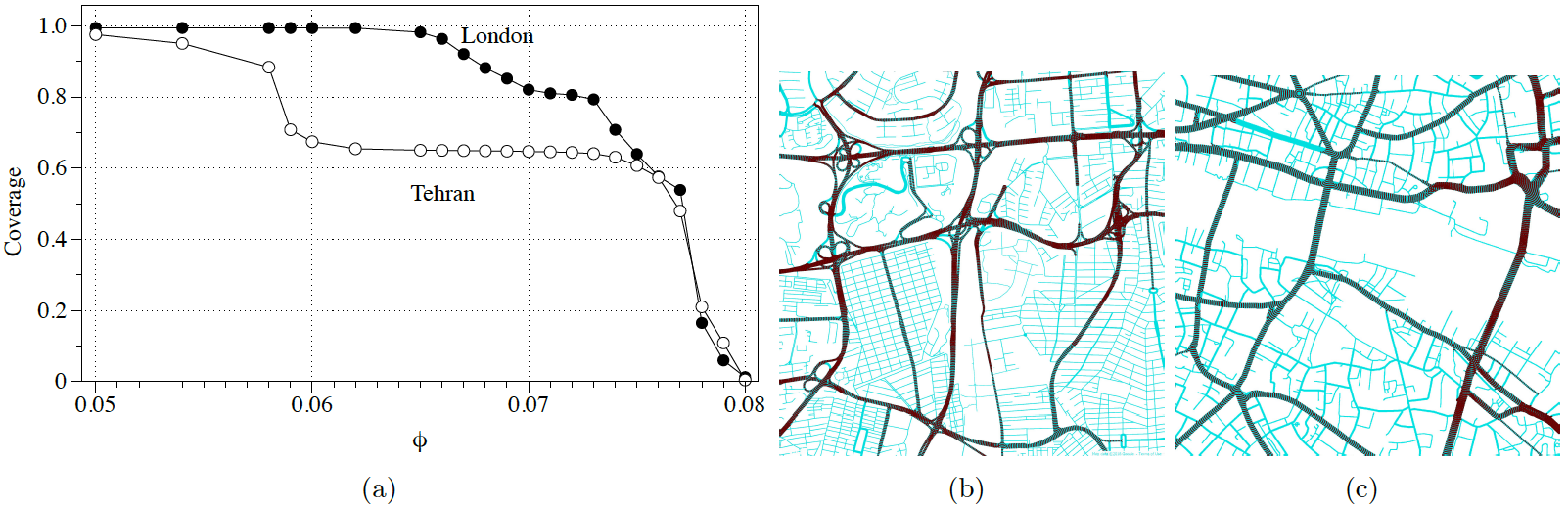}
    \caption{Excitable media based comparison of Tehran and London. (a)~ Coverage of Tehran and London~\cite{adamatzkyExcitableLondon} street network fragments by excitation wave front depending on excitability $\phi$. (bc)~Propagation of excitation on the street map of Tehran~(b) and London(c). These are time lapsed snapshots of a single wave-fragment recorded every 150\textsuperscript{th} step of numerical integration. Maps are generated using Processing www.processing.org).  Videos, snapshots and data files are available at DOI 10.5281/zenodo.1304036.}
    \label{fig:TehranLondon}
\end{figure}

There are noticeable differences in coverage of a selected fragment of Tehran street network and a fragment of London street network~\cite{adamatzkyExcitableLondon}, see Figure~\ref{fig:TehranLondon}. Phases $P_1$ and $P_2$ and the transitions between them are present on the coverage vs. $\phi$ plot of London, however they are less pronounced than that of Tehran. In the case of London, the phase $P_1$ lasts till $\phi=0.065$ with coverage nearly 1. This may be explained by the fact that on the fragment of Tehran street network there is a plenty of narrow streets, branching at straight angles from the wider, major, streets. Transition period $T_1$ in London lasts till $\phi=0.07$. The phase $P_2$ is relatively short, from $\phi=0.07$ to $0.073$. Excitability value $\phi=0.076$ is shown (Figure~\ref{fig:TehranLondon}a) to be a critical one, for this value of $\phi$ coverage of Tehran (Figure~\ref{fig:TehranLondon}b) and London (Figure~\ref{fig:TehranLondon}c) street networks converge. Space-time dynamics of excitation well reflect differences in geometry of street networks of two cities studied, however, to make any further generalisations we must undertake a set of comparative experiments on a larger pool of street networks. That will be a scope of further studies.

\begin{figure}[!tbp] 
    \centering
      \includegraphics[width=\textwidth]{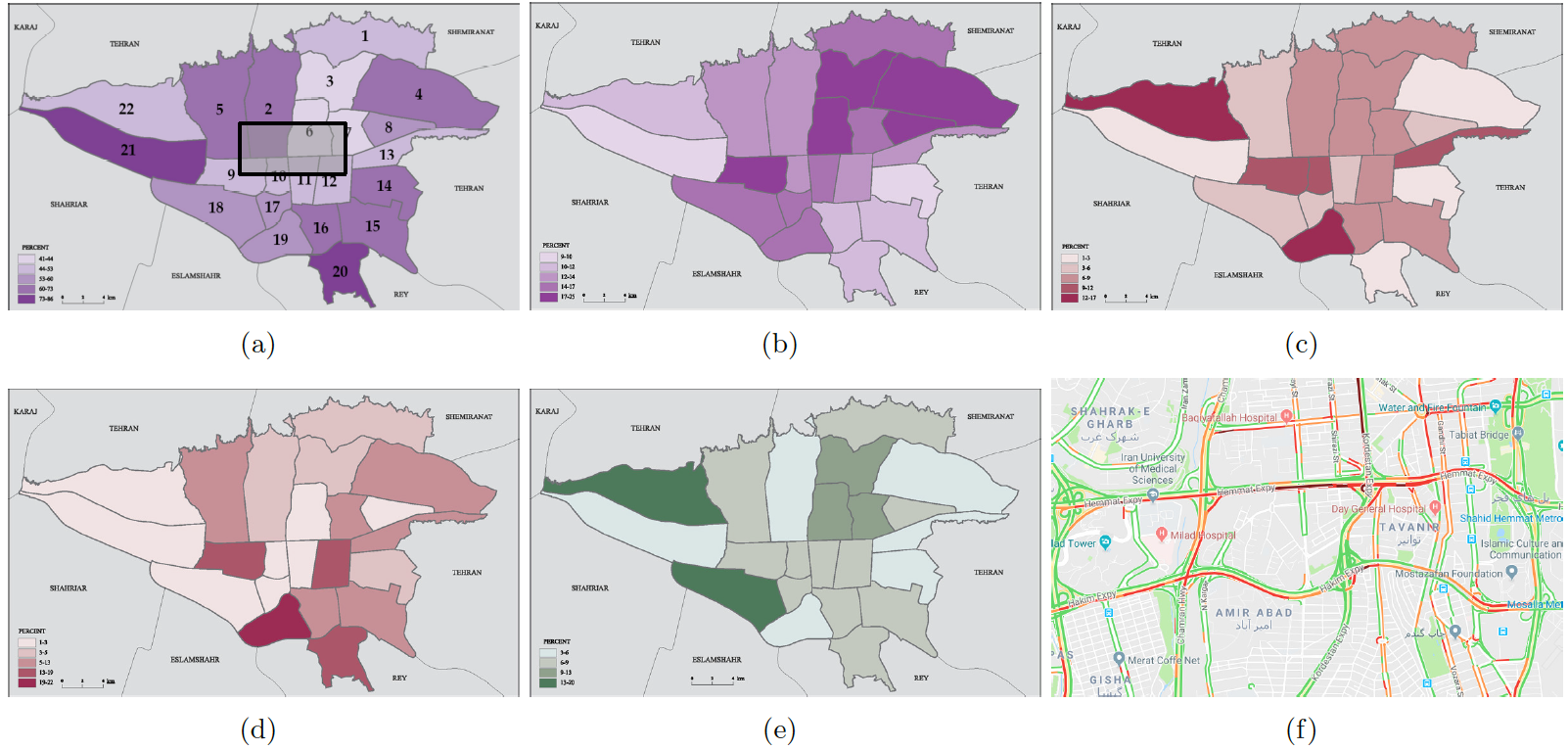}
    \caption{Realities of Tehran street network (ae)~Infographic of $\zeta$ ratio in Tehran districts. 
    (a) Free flow traffic condition. Fragment of Tehran street network, studied by us, is shown by grey rectangle. 
    (b) Moving traffic condition. 
    (c) Slow traffic condition. 
    (d) Congested traffic condition. 
    (e) Acceptable traffic condition. 
    \copyright Tehran Municipality.  Tehran municipality, Public \& International Relations Department.
    (g)~Live traffic of the selected region \copyright 2018 Google map.}
    \label{fig:TehranTraffic}
\end{figure}
\graphicspath{{figs/}}

To highlight the importance of the results and hint on their relevance to traffic models, let us substantiate our choice of the street network fragment. The road network in Tehran is evaluated by Tehran Municipality by a $\zeta=V/C$ ratio, where $V$ is a total number of vehicles passing a point in one hour and $C$ is the maximum number of cars that can pass a certain point at the reasonable traffic condition: free traffic $\zeta<0.7$, moving traffic $0.6 \leq \zeta \leq 0.9$, moderate traffic $0.9 < \zeta \leq 1.1$, heavy traffic $1.1 < \zeta \leq 3.1$, congested traffic $3.1 < \zeta$. Based on $\zeta$, the traffic flow in 21 municipal districts of Tehran can be categorised as follows (Figure~\ref{fig:TehranTraffic}). Districts 21 and 15 have moving traffic conditions. Districts 4, 6, and 10 have moderate traffic flows, districts 2, 22 and 18 have heavy traffic condition, districts 17, 12, 11, 10, 6, 8, and 7 almost experience congested traffic condition. Twenty-two districts of the Tehran experience a total of nearly fourteen million daily vehicular trips in which the district 4 is the highest origin of trips, followed by districts 15, 2, and 5. Districts 12 and 6 have the highest number of trips destinations. Specifically, the greatest number of educational trips are made between District 4, as the origin, and District 6, as the destination destinations. While the shopping trips have origins in Districts 4, 2, and 15 and destined in districts 6 and 12.

The region selected  in our studies lies in districts 2, 6, 7, 10, 11, and 12 (Figure~\ref{fig:TehranTraffic}f) because its neighbouring districts show substantial variety in traffic conditions (Figure~\ref{fig:TehranTraffic}). This region contains main highways linking the east-west and north-south of Tehran which cross each other. The majority of traffic among different districts go thorough Hakim, Hemmat, Yadegar-E-Imam, and Modarres highways. Indeed, if local ways face heavy or congested traffic, this traffic will be propagated to these key highways (Figure~\ref{fig:TehranTraffic}g). With regards to the traffic the following observations could be explored in more details in further studies: (1)~higher traffic peak, in reality, might correspond to faster movement of excitation wave-fronts, (2)~increasing value $\phi$ might show unpredictability of the travel, e.g. in a rainy day traversing house increases exponentially, (3) dynamics of excitation for $\phi=0.08$ reflects congestion when Hemmat path reaches Hakim. 

Evaluating street networks in terms of earthquake vulnerability might be another application domain for excitable media. To minimise earthquake damages, it is useful to estimate traffic patterns and accessibility of a city after an earthquake~\cite{modarres2002application}; in ~\cite{modarres2002application}; the city street network is evaluated using the criterion of accessibility based on travel time and safety. Assuming an earthquake damage is less pronounced at wider streets, we could propose that excitability value $\phi$ characterises accessibility: excitable medium with higher values $\phi$ select streets which could be accessible after an earthquake.

\bibliographystyle{plain}
\bibliography{bibliography}

\end{document}